\documentstyle[aps,epsf,pre,multicol,floats]{revtex}

\begin{document}
\title{Diblock Copolymer Ordering Induced by Patterned Surfaces Above
  the Order-Disorder Transition}
\author{Yoav Tsori and David Andelman$^*$} 
\address{School of Physics and
  Astronomy,
  Raymond and Beverly Sackler Faculty of Exact Sciences\\
  Tel Aviv University, 69978 Ramat Aviv, Israel} 
\date{3/7/00}
\maketitle

\begin{abstract}

We investigate the morphology of diblock copolymers in the
vicinity of flat, chemically patterned surfaces. 
Using a Ginzburg-Landau free energy, spatial
variations 
of the order parameter are given in terms of a general
two-dimensional surface pattern above the
order-disorder transition. The propagation of several
surface patterns into the bulk is investigated. The oscillation
period and decay length of the surface $q$-modes are calculated in
terms of system parameters. We observe lateral order
parallel to the surface as a result of order perpendicular to
the surface. Surfaces which has a finite size chemical pattern
(e.g., a stripe of finite width) induces lamellar ordering
extending into the bulk. Close to the surface pattern the lamellae are
strongly perturbed adjusting to the pattern.

\end{abstract}

\pacs{PACS numbers 61.25.Hq, 83.70.Hq, 61.41.+e, 02.30.Jr}

\begin{multicols}{2}
\section{Introduction}\label{intro}

The bulk properties of diblock copolymers (BCP) are now well
understood \cite{B-F90,O-K86,Leibler80,F-H87,M-B96}. These long linear
macromolecules composed of two incompatible sub-chains, or blocks, cannot 
phase separate because of the covalent bond between them. This
connectivity, together with the incompatibility between the two
blocks, gives rise to the appearance of microphase separated phases.
The state of segregation is controlled by $N\chi$ and $f$, where
$\chi$ is the Flory parameter, $N=N_A+N_B$ is the total number of
constituents monomers per chain and $f$ is the fraction of the $A$
block, $f=N_A/\left(N_A+N_B\right)$. For large enough $N\chi $ one of
the ordered phases, such as the lamellar, hexagonal or cubic phases is
preferred, depending on the degree of asymmetry $f$.

Less understood is the interfacial behavior of copolymer melts near solid
surfaces or at the free surface with air. Surface phenomena of BCP may
enable creating and controlling technologically important devices 
of characteristic size comparable to the wavelength of light.
As examples we mention
waveguides, light-emitting diodes and other optoelectronic device,
anti-reflection coating for optical surfaces
\cite{M-SSCIENCE99} and dielectric mirrors \cite{fink98}.

The presence of a wall in a BCP system
leads to new energy and length scales, depending on the specific chemical
interaction of the polymers with the surface. In a semi-infinite system in
contact with a single planar wall, the
morphology near the surface can be very different from the bulk morphology.
Fredrickson \cite{Fredrickson87} has considered BCP in contact with a
surface having a uniform preferential adsorption 
to one of the two blocks. Above the
order-disorder
transition (ODT), where $\chi<\chi_c$ ($\chi_c$ is the critical point value of
$\chi$ above which an ordered phase appears), 
he used mean-field theory
and found that the order parameter (being the concentration difference between
the two blocks) has decaying oscillations. He showed that the oscillation
periodicity depends on $\chi$, and tends to the bulk lamellar periodicity as
$\chi \rightarrow \chi_c$. In the same $\chi\rightarrow \chi_c$
limit, the correlation length $\xi$ of the oscillations was found to diverge.
Further investigations \cite{T-F92} showed that the inclusion of higher order,
nonlinear corrections to the mean-field theory results in a non-diverging
$\xi$. For the same system cooled below the ODT, modulated sinusoidal behavior
was found.
In a related work \cite{T-R-M94} a Ginzburg-Landau free energy was used to
describe the propagation
 of a surface-induced lamellar
ordering into a bulk hexagonal phase. In the strong-segregation limit
a lamellar region of finite thickness close to the surface becomes
stable, provided that the surface field is larger than some critical value.

The situation is even more complicated in thin films, where the
distance between the two boundaries, associated with the film
thickness, is comparable to the periodicity of modulations in the
bulk, and the surface induced morphology can be of different symmetry
than that of the bulk. For a system taken in one of its ordered phases (below
the ODT), the free energy has a local minimum when the spacing between
the surfaces is an integer multiple of the bulk repeat period. The
mean-field behavior of BCP close to surfaces and for BCP films was
calculated \cite{shull92} using a method applicable in both the strong
and weak segregation limits. It was found that confinement of
lamellar phase BCP may lead to parallel layering, or in some cases
even to a perpendicular arrangement of the lamellae. Self-consistent
field theory (SCF) was used \cite{matsenJCP97,G-M-B00} to study the
stability of these parallel, perpendicular and mixed lamellar phases
in thin films of BCP. The latter phase consists of parallel lamellae
near one surface and perpendicular lamellae near the opposite surface,
but it was found to be unstable for symmetric A-B ($f=1/2$) diblock
copolymers.

So far, we mentioned situations where the surfaces have a uniform
preference to one of the two blocks. More complex, chemically
patterned surfaces break the lateral translation symmetry.  Different
surface regions will now have a different preference for the A/B
blocks, thereby inducing a lateral structured morphology near 
the surface.
Very few works took into account this possibility of a non-uniform
surface. In particular,
Petera and Muthukumar \cite{P-Muthu97,P-Muthu98} have
investigated the effect of a {\it one dimensional} sinusoidal surface
pattern on BCP 
morphologies close to the surface in the weak-segregation limit, both
below and above the ODT. 

In this paper we consider a BCP melt above the ODT near a surface,
whose pattern is truly arbitrary in two dimensions, generalizing the
results of Refs.
\cite{P-Muthu97,P-Muthu98}. 
A Ginzburg-Landau free energy is expressed
in term of the polymer concentration  is presented
in Sec. \ref{model}. In Sec. \ref{1d} we consider a melt close to one
surface or confined
between two surfaces whose chemical pattern has one
dimensional symmetry. Minimization of the free energy expansion gives
rise to an Euler-Lagrange equation for the order parameter. A natural
generalization to two-dimensional surface patterns is then considered
in Sec. \ref{2d}. We are able to give a complete description of the
order parameter in terms of all the $q$--modes of the surface pattern.
Finally, conclusions
and some future prospects are presented in Sec.  \ref{conclusions}.

\section{The model}\label{model}

The copolymer melt is described by the order parameter $\phi({\bf
  r})$, defined as $\phi({\bf r})=\phi_A({\bf r})-f$, the difference
in local A monomer concentration from its average value. 
Hereafter we restrict the treatment to the symmetric $f=1/2$ case, following 
the same coarse-grained free energy as was used by
Fredrickson and Binder \cite{Binder_Fredrickson89,B-F-SJPII97}:

\begin{equation}\label{F}
\frac{N}{k_BT}F=\int\left\{\frac12\phi\left[\tau+h\left(\nabla^2+q_0^2\right)^2\right]\phi +\frac{u}{4!}\phi^4\right\}{\rm d}^3{\bf r}
\end{equation}
Where $k_B$ is the Boltzmann constant and $T$ is the temperature. The
other parameters are:
\begin{eqnarray}
q_0\approx 1.9456/\sqrt{\langle R_g^2\rangle}\\
\tau=2\rho N\left(\chi_c-\chi\right)\\
\chi_c=10.495/N\\
h=1.5\rho c^2\langle R_g^2\rangle/q_0^2
\end{eqnarray}
The fundamental wavelength of the system, $q_0$, is expressed by
$R_g$, the radius of gyration of the chains. The chain density
$\rho$ is equal 
to $1/Na^3$ for an incompressible melt, and $u/\rho$ and $c$ are of order
unity. More details can be found in Ref.~\cite{B-F-SJPII97} and extensions
for asymmetric BCP, $f\neq 1/2$ are possible as well. 
The use of (\ref{F}) limits our treatment to a region of the phase diagram
close enough to the critical point where the expansion in powers of
$\phi$ and its derivatives is valid, but
not too close to it, because then critical fluctuation effects may be important
\cite{brazovskii}.

This and similar types of free energy has been used to describe bulk
and surface phenomena in amphiphilic systems \cite{G-S90}, diblock
copolymers \cite{Leibler80,F-H87,B-F-SJPII97,T-A-S00,N-A-SPRL97},
Langmuir films \cite{A-B-J87} and magnetic (garnet) films
\cite{G-D82}. The $\phi^2$ and $\phi^4$ terms appear in the usual
Landau expansion. The added $\phi\nabla^2\phi$ 
and $\phi\nabla^2\nabla^2\phi$ terms
compete to produce {\em modulated phases} below the order-disorder
temperature. This free energy describes a system in the
disordered phase ($\phi=0$, $f=1/2$) for $\chi<\chi_c$, and in
the lamellar phase for $\chi>\chi_c$. The $q=q_0$ mode goes critical
first, and the lamellar phase is described by $\phi=\phi_q\cos({\bf
  q_0\cdot r})$, of repeat period $d_0\equiv 2\pi/q_0$. This
single-mode approximation is accurate to order $(\chi-\chi_c)^{1/2}$
and can be justified near the critical point \cite{Fredrickson87}. Far
from the critical point higher harmonics are needed to describe the
lamellar phase.  As the asymmetry in composition is increased, other
ordered phases of hexagonal and cubic symmetries become more stable
than the lamellar phase.

As stated above, block copolymers exhibit complex surface behavior
characterized by
the strength and range of the
interaction between the polymer chains and the surface, the typical size
of chemical heterogeneities of the surface, 
and the distance between the two
surfaces, in case of a thin film.

The presence of chemically interacting confining walls is modeled by an
added short-range surface coupling term in the free energy,
\begin{equation}\label{Fs}
F_s=\int{\rm d^2{\bf r_s}}\left(\sigma({\bf r_s})\phi({\bf r_s}) +\tau_s\phi^2({\bf
r_s})\right)
\end{equation}
The vector ${\bf r_s}$ define the position of the confining surfaces.
The $\sigma\phi$ term expresses the preferential interaction of the
surface with the A and B blocks. For example, if $\sigma>0$ then the B
block ($\phi<0$)is attracted to the surface more than the A block
($\phi>0$).
  Control over
the specificity of this surface term can be achieved by coating the
substrate with carefully prepared random copolymers
\cite{L-RPRL96,M-RPRL97}. The coefficient of the $\phi^2$ term in
(\ref{Fs}), $\tau_s$, is a surface correction to the Flory parameter $\chi$
\cite{Fredrickson87,T-F92,K-Muthu99}. $\tau_s>0$ corresponds to a
suppression of surface segregation of the A and B monomers.

We first consider 
systems in which the polymer melt is confined by a flat,
rigid wall at $y=0$, with the $x$-axis chosen in the plane of the
wall, and is translational invariant along the $z$-direction.
Extension to the system of two parallel surfaces located at $y=\pm L$
is straightforward and will be
considered later. The order parameter $\phi$ is expected to vanish in
regions where the interfacial interactions can be neglected,
\begin{equation}\label{bcphi}
\lim_{y\rightarrow\infty}\phi(x,y)=0
\end{equation}
recovering the value $\phi=0$ of the bulk phase far from the surface.
In the next section we find profile solutions $\phi(x,y)$ for a
BCP system at temperatures above the bulk ODT.

\section{One dimensional surface patterns}\label{1d}

For high enough temperatures, or equivalently, for $\chi<\chi_c$, the
phase of lowest free energy is the homogeneous disordered phase, with
$\phi({\bf r})=0$ 
in the bulk. The presence of a patterned surface induces ordering
in the copolymer 
melt. If the chemical surface composition is uniform, one of the
monomers, A or B, will be attracted to the surface, resulting in a parallel
orientation of the lamellae (a perpendicular orientation of the
chains). If the pattern is modulated, say sinusoidally, then different
blocks are attracted to different regions of the surface.  
We will start with this case
of a semi-infinite system bounded by one rigid, flat surface, and then
proceed to describe thin film systems between two surfaces.

\subsection{One patterned surface}\label{1surface}

Consider the semi-infinite BCP melt at $y>0$ bounded by a flat surface
given by $y=0$.  We assume a one dimensional periodic surface pattern
and write it in terms of the Fourier components of the surface field
$\sigma(x)$

\begin{equation}\label{sigma}
\sigma(x)=\sum_q\sigma_{q}{\rm e}^{iqx}
\end{equation}
where $\sigma_q$ set the amplitude of the respective
$q$--modes. The order parameter $\phi(x,y)$ satisfies the
boundary conditions on the surface and approaches the bulk solution
far from the surface. It is
convenient to decompose $\phi$ in terms of its $q$--modes in the
$x$-direction
\begin{equation}\label{ansatz}
\phi(x,y) =\sum_qf_q(y) {\rm e}^{iqx}
\end{equation}
The requirement (\ref{bcphi}) leads to the bulk boundary condition
\begin{equation}\label{bcg}
\lim_{y\rightarrow\infty}f_q(y)=0
\end{equation}
The form (\ref{ansatz}) is substituted in (\ref{F}). Above the
order-disorder transition (ODT) temperature, the theory is stable to
second order in $\phi$, and therefore the $\phi^4$ term is neglected.
Using the explicit $x$-dependence of $\phi$ in (\ref{ansatz}) we
perform the $x$ and $z$ integration, yielding the free energy $F$:
\begin{eqnarray}\label{avg-Df}
F&=&\int\sum_q\left\{\left(\tau+hq_0^4\right)f_qf_q^*+hq_0^2\left[f_q\left(f_q^{''}-q^2f_q\right)^*+c.c.\right]\right.\nonumber\\
&+&\left.\frac12h\left[f_q\left(f_q^{''''}-2q^2f_q^{''}+q^4f_q\right)^*+c.c.\right]\right\}{\rm dy}\nonumber\\
&+&\sum_q\left(\sigma_qf_q^*(0)+\tau_sf_q(0)f_q^*(0)\right)+ c.c.
\end{eqnarray}
where $(...)^*$ indicates complex conjugation ( $c.c.$ ) operation.
A standard minimization technique is carried on and yields the
governing linear ordinary differential equation for the functions
$\{f_q\}$ for $y>0$:

\begin{equation}\label{EL-gq}
\left(\tau/h+\left(q^2-q_0^2\right)^2\right)f_q+2(q_0^2-q^2)f_q^{''}
+f_q^{''''}=0\label{gov-gq}
\end{equation}
This equation possesses four independent solutions in the form of an
exponential ${\rm e}^{-k_qy}$, with $k_q$ found from the characteristic
equation
\begin{equation}\label{kq-eqn}
\left(\tau/h+\left(q^2-q_0^2\right)^2\right)+2(q_0^2-q^2)k_q^2+k_q^4=0
\end{equation}
Thus, with the semi-infinite geometry, the solution is
\begin{equation}
f_q(y)=A_q {\rm e}^{-k_qy} + B_q{\rm e}^{-k_q^*y}\label{gq}
\end{equation}
and
\begin{equation}
k_q^2 =q^2-q_0^2+ i\left(\tau/h\right)^{1/2}\label{kq}
\end{equation}
Each $q$-mode solution $f_q$ is characterized by two complex amplitudes
$\{A_q,B_q\}$. From the solutions of Eq. (\ref{kq-eqn})
wavevectors $\{k_q\}$
with negative real value, ${\rm Re}(k_q)<0$, are discarded, to comply
with the boundary condition (\ref{bcg}). Note that ${\rm Re}(k_q)$ is
increasing monotonously as a function of $q$. A large value of ${\rm
  Re}(k_q)$ means short decay length, and hence the smallest surface
$q$--mode decays the least. This behavior is demonstrated on Fig.~1
showing the $q$ and $\chi$ dependence 
of the real and imaginary parts of the wave-vector $k_q$. For a fixed
value of $\chi$ the decay length, proportional to $1/{\rm Re}(k_q)$,
decreases as $q$ increases, while the wavelength $2\pi/{\rm Im}(k_q)$
of the modulations in $f_q(y)\sim {\rm e}^{-k_qy}$ increases.

The boundary conditions for the functions ${f_q}$ can be determined by
considering the Euler-Lagrange equation for $\{f_q\}$ in the range
that includes $y=0$. In this case a term proportional to the Dirac delta
function $\delta(y)$ appears in (\ref{EL-gq}). There are two
conditions relating $f_q$ and its derivatives at $y=0$:
\begin{eqnarray}\label{bcgq1}
f_q^{''}(0)+2\left(q_0^2-q^2\right)f_q(0)&=&0\\
\frac{2\sigma_q}{h}+\frac{4\tau_s}{h}f_q(0)+2(q_0^2-q^2)f_q^{'}(0)+f_q^{''}(0)
&=&0\label{bcgq3}
\end{eqnarray}

Recently, it has been found \cite{mukamelPRE00} that surface states
exist even in the absence of a surface field $\sigma$. This effect can
be attributed to a loss of entropy close to the surfaces. However, in
our linear theory this does not happen, and the copolymer response is
proportional to the surface field $\sigma$.
The case where $\sigma_0$ is a non-zero constant and $\sigma_{q\neq 0}=0$,
corresponds to the special case of uniform interfacial interactions.  The
system exhibits a decaying lamellar layering of the polymers, with the
B--polymer adsorbed to the surface if $\sigma_0>0$. 

Close to the ODT,
the complex wavevector $k_0$ can be approximated by
\begin{equation}
k_0\simeq -\frac{\left(\tau/h\right)^{1/2}}{2q_0}+iq_0\left(1-\frac{\tau/h}{8q_0^4}\right)
\end{equation}
This expression shows that $f_0\sim {\rm e}^{-k_0y}\sim {\rm e}^{-y/\xi}$ has a
diverging characteristic length
$\xi\propto\left(\chi_c-\chi\right)^{-\frac12}$, while the oscillatory
part has a wavelength slightly longer than that of the bulk lamellar
phase, in agreement with the results of Fredrickson for chemically
uniform surfaces \cite{Fredrickson87}. The correlation length $\xi$
diverges for small composition oscillations because a linear
theory is employed; addition of the $\phi^4$ term in $F$ would give a
finite value 
of $\xi$.  As a result of the assumed short-range surface
interactions, the periodicity and decay length of $f_q$
depend only on properties of the bulk, and not on surface details.

Using the notation $k_q=k_q^{'}+ik_q^{''}$ the real part of 
the form (\ref{gq}) can be
rewritten as: 
\begin{eqnarray}\label{alpha}
2{\rm Re}\left(f_q\right)=\left(A_q+B_q^*\right){\rm e}^{-k_qy}
+c.c.\nonumber\\
=2|A_q+B_q| {\rm e}^{-k_q^{'}y}\cos(k_q^{''}y+\alpha_q)
\end{eqnarray}
where $\alpha_q$ is the phase of the $q$-mode modulation. It determines the 
value of the $f_q$ solution at the boundary, $y=0$.
 
The phase $\alpha_0$ for the $q=0$ mode is found to satisfy the
following relation
\begin{equation}
\tan\alpha_0=\frac{q_0^2}{\sqrt{\tau/h}}
\end{equation}
and therefore is determined by the degree of segregation $\chi$,
but not by the pattern amplitude $\sigma_q$. A plot of $\alpha_q$ as a
function of $q$ for several values of $\chi$ is shown in Fig.~2 (a).
Far from the ODT point and deep into the disordered phase, $\chi\ll\chi_c$,
 we find that all $q$-modes have
a phase angle $\alpha_q=0$. As the ODT is approached, the $q=q_0$ mode
retains its value but larger $q$-modes have a negative phase while smaller
$q$-modes have a positive phase. At $\chi=\chi_c$ this becomes a step
function, with $\alpha_q=\pi/2$ for $q<q_0$ and
$\alpha_q=-\pi/2$ for
$q>q_0$. The amplitude behavior is shown in Fig.~2 (b), for the same
series of segregation values $\chi$. As the ODT is approached, the
$q=q_0$ mode becomes critical first, with a diverging amplitude.

An interesting limit occurs when $\sigma_0=0$, that is, the average
surface interaction is zero (no net adsorption).  No lamellar ordering
parallel to the surface is expected. Indeed, the resulting
checkerboard behavior is
illustrated for a surface pattern chosen for simplicity to contain
only one mode: $\sigma(x)=\sigma_q\cos(qx)$.  Fig.~3 depicts
alternating A-rich (white) and B-rich (black) regions. In (a) the
decay length $\xi$ is smaller than in (b), because in the former case
the surface periodicity is twice as large.
The
oscillatory behavior, characterized by ${\rm Im}(k_q)$, has a very
long wavelength, diverging as $(\chi_c-\chi)^{-1/2}$ close to the ODT
point.

Usually, if no special measures are taken \cite{L-RPRL96,M-RPRL97},
there is a net preference to one of the monomers: $\sigma_0\neq 0$.
The BCP morphology where the surface interactions were chosen to have
both a non-zero average preference and undulatory character, namely
$\sigma=\sigma_0+\sigma_q\cos(qx)$, is shown in Fig.~4.  A smooth
crossover from surface-induced ordering at small distance to the bulk
disorder occurs.
The parallel lamellae resulting from the $\sigma_0$ term
persist farther from the surface than the bulges resulting from the
$\sigma_q$ term, as $f_0$ decays slower than $f_q$. For a given
$\sigma_0$, having a higher $q$-mode or reducing the modulation
strength $\sigma_q$ will enhance the lamellar features far from the
surface.

\subsection{Two patterned surfaces}\label{2surface}

Until now the BCP melt was assumed to be bounded by one
surface at $y=0$. In this section we extend our analysis to a
thin-film system confined between two parallel surfaces
located at $y=L$ and $y=-L$, shown in Fig.~5. 
When the distance $2L$ between the surfaces is
comparable to the natural bulk periodicity, the two surfaces interact
via the BCP and
the resulting film morphology can be very 
different from that of the one-surface case (Sec. \ref{1surface}). However,
the mathematical 
analysis is almost the same; one only has to apply different boundary
conditions on the BCP order parameter $\phi$. 

The surfaces at $y=\pm L$ are assumed to carry different surface
fields of the form $\sigma^{\pm}(x)=\sum_q\sigma_q^{\pm}{\rm e}^{iqx}$.
Only small modifications must be included to adjust the results of the
previous section. The same ansatz (\ref{ansatz}) for
$\delta\phi$ is used.
The functions $f_q$ that minimize the appropriate $x$-averaged free energy
(\ref{F}) obey
\begin{equation}\label{gq1}
f_q(y)=A_q {\rm e}^{-k_qy}+B_q {\rm e}^{-k^*_qy}+C_q {\rm e}^{k_qy}+D_q 
{\rm e}^{k^*_qy}
\end{equation}
with $\{k_q\}$ given by the same relation (\ref{kq}). However, unlike the
semi-infinite bulk one-surface case (\ref{gq}),
both
signs of the $k$ vectors are used because the system is finite in the
$y$-direction. In 
addition, repeating the procedure outlined for the one-surface case, 
gives the boundary conditions for $f_q$:
\begin{eqnarray}\label{bcg02}
f_q^{\prime\prime}(\pm L)+2\left(q_0^2-q^2\right)f_q(\pm L)&=&0\\
\frac{2\sigma_q^{\pm}}{h}+\frac{4\tau_s}{h}f_q(\pm L)\\
\mp 2\left(q_0^2-q^2\right)f_q^{\prime}(\pm L)\mp f_q^{\prime
\prime\prime}(\pm L)&=&0\nonumber
\end{eqnarray}

We consider now several specific surfaces. In the first $\sigma^+=1$
is a constant, while $\sigma^-=\cos(qx)$ is purely sinusoidal and average to
zero, as depicted in
Fig.~6. As expected, the B polymer (in black) is attracted to the
upper surface, while the bottom surface exhibits modulated
adsorption pattern. Although lamellar features are seen near the
top surface, the overall apparent phase in the sample cannot be
classified as such. The corresponding plots of the functions
$f_0(x)$ and $f_q(x)$ are shown in Fig.~7 (same parameters as in
Fig.~6). In general $f_q$ is nonzero even at the $y=L$ surface,
although the surface does not induce modulations by itself, $\sigma^+=const$. 
Thus
modulations propagate from one surface to the other by the
copolymer melt. This is an interesting observation which may
have relevance in applications. It relies on the relative small
thickness of the BCP film.

The situation as depicted in Fig.~6 represents a competition
between two mechanisms. The modulated pattern at the bottom surface
induces a laterally modulated pattern of the BCP, while the
top surface uniform interaction induces a lamellar-like layering of
the copolymers. As the modulated adsorption pattern strongly depends on
the modulation wavenumber, so does the resulting morphology. This
effect is shown explicitly in Fig.~8, where the top surface is
uniform and the bottom is modulated, for a series of $q/q_0$
values. The transition from a locally perpendicular (bottom
patterned surface) to a locally parallel orientation (at the top
uniform surface) is seen in (a), similar to the so-called
T-junctions between grains of different orientations
\cite{N-A-SPRL97,G-TMAC94}. Similar behavior was found by the SCF
calculation in Ref. \cite{P-Muthu98}.

Figure.~9 (a) shows the spatial dependence of the BCP order parameter when the
two surfaces contain only one $q$-mode and are patterned {\it in
phase} with each other (symmetric
arrangement), but with opposite signs, $\sigma^{\pm}=\pm
\sigma_q\cos(qx)$. The 
copolymer patterns create a perfect checkerboard arrangement and are related to
each other at the surfaces by an interchange of monomers A$\leftrightarrow$B.
The surface pattern (\ref{sigma}) contains only $\cos(qx)$ terms. A
generalization that includes $\sin(qx)$ sinusoidally varying modes is
straightforward. In this case the patterns at the surfaces can be
out--of--phase with each other. Figure.~9 (b) shows such a morphology, for
$\sigma^+=\sigma_q\cos(qx)$, $\sigma^-=\sigma_q\sin(qx)$, where there is a
$\pi/2$ phase shift between the two surface fields. The perfect checkerboard
arrangement of 9 (a) is now distorted to accommodate this phase shift.

\section{Two-dimensional surface patterns}\label{2d}

So far we considered a melt in contact with a surface or confined
between two surfaces of one
dimensional symmetry. In our approximation,
like in any linear
response theory, there is no $q$-mode coupling proportional to
$\sigma_{q_1}\sigma_{q_2}$. This fact allows us to go further and
introduce a two-dimensional generalization of the surface pattern,
which so far was taken to be independent on $z$. The surface now is
assumed to carry a chemical pattern $\sigma(x,z)$ which can be written
as:
\begin{equation}
\sigma(x,z)=\sum_{q_x,q_z}\sigma_{q_x,q_z} {\rm e}^{i\left(q_xx+q_zz\right)}
\end{equation}
The ``linear response'' function is then
\begin{equation}
\delta\phi(x,y,z)=\sum_{q_x,q_z}f_{q_x,q_z}(y) {\rm e}^{i\left(q_xx+q_zz\right)}
\end{equation}
Because $f$ and $\sigma$ are real functions,
$f_{-q_x,-q_z}=f^*_{q_x,q_z}$ and similarly for $\sigma$. Using the
above form it is possible to carry out the integration of the free
energy in the $x-z$ plane. Denoting $\langle . . .\rangle_{xz}$ as the
average in the $x-z$ plane, it can be checked, for example, that
\begin{equation}
\langle \phi\nabla^2\phi\rangle_{xz}=
\sum_{q_x,q_z}f_{q_x,q_z}\left(f_{q_x,q_z}^{\prime\prime}
-\left(q_x^2+q_z^2\right)f_{q_x,q_z}\right)^*
\end{equation}
Defining ${\bf q_{_{\parallel}}}\equiv \left(q_x,q_z\right)$ and
performing the free energy minimization with respect to $f_{q_x,q_z}^*$,
the functions $f_{q_{\parallel}}=f_{q_x,q_z}$ obey the same master
equation (\ref{gov-gq}) that $f_q$ previously obeyed, with the only
change that $q^2$ is replaced by $q_{_{\parallel}}^2$. For a BCP in contact
with a single surface, the appropriate boundary conditions are:
\begin{eqnarray}
f_{q_{\parallel}}^{\prime\prime}(0)+2\left(q_0^2-
q_{_{\parallel}}^2\right)f_{q_{\parallel}}(0)&=&0\nonumber\\
\frac{2\sigma_{q_{\parallel}}}{h}+\frac{4\tau_s}{h}f_{q_{\parallel}}(0)+
+2\left(q_0^2-q_{\parallel}^2\right)f_{q_{\parallel}}^{\prime}(0)+f_{q_{\parallel}}^{\prime\prime\prime}(0)&=&0
\end{eqnarray}
The solution for $f_{q_{\parallel}}$ is analogous to (\ref{gq}),
\begin{equation}
f_{q_{\parallel}}(y)=A_{q_{\parallel}}{\rm e}^{-k_{q_{\parallel}}y}+
B_{q_{\parallel}}{\rm e}^{-k^*_{q_{\parallel}}y}\label{mod-gq}
\end{equation}
\begin{equation}
k_{q_{\parallel}}^2 =q_{_{\parallel}}^2-\frac 12+
i\left(\chi_c-\chi\right)^{1/2}\label{mod-kq}
\end{equation}
Having found the response of the polymers to the surface modes
$\sigma_{q_{\parallel}}$, one is able to deduce the concentration
profiles for any given two-dimensional surface pattern. In order to
illustrate this, we take a system of chemical affinity in the shape of
the letters ``BCP'' on the $y=0$ surface [Fig.~10 (a)], and calculate
the polymer concentration in the planes parallel and above it. All
sizes are expressed in terms of $d_0$, the lamellar fundamental periodicity. 
The
shape of the letters continuously deforms as one moves away from the
surface. Contour plots corresponding to planes parallel to the $x-z$ surface
and separated by a
distance $(n+1/2)d_0$, for integer $n$, are approximately given by an
$A\leftrightarrow B$ interchange of monomers. This is the
characteristic distance at which the polymers flip.  Note that Fig.~10
(c) and (e) are approximately the inverse image (``negative'') of (b),
(d) and (f). The 
original features are completely washed away as the distance $y$ from
the surface is further increased. In our case for $5d_0\lesssim
y\lesssim 6d_0$ where the
surface pattern size was roughly $d_0$. 

Figure 10 also illustrates
circumstances where a certain surface pattern is transferred via
the bulk BCP to another, distant surface. It may be important to know, for
example, if the contrast of the distant image can be experimentally
detected. This reduction of the contrast is clearly seen by comparing Fig.~10
(b) and (f), and can easily be calculated from our expressions. The
lamellar order created parallel to the edges of the letters in Fig.~10 (b) is
the result of the undulatory nature of the block copolymers: order
extending perpendicular to the surface induces order in the direction
parallel to it.

The copolymer melt can follow the surface pattern when its size is
larger than the polymer length-scale $d_0$. The effect of
reducing the size of the surface structure is seen as a blurred
morphology in Fig.~11 (a), where the ``BCP'' pattern was chosen to
have dimensions $4d_0\times 4d_0$, compare to $20d_0\times 20d_0$ of
Fig.~10 (b).  The effect 
of raising the temperature (further away from the ODT) 
is seen in Fig.~11 (b). It is similar
to Fig.~10 (b), only that the temperature is higher, $N\chi =9$, and the
lamellar features along the edges of the letter are less prominent.
Again, using our order parameter expressions one can quantify the
$q$-mode spectrum and contrast of the distant image, as a function
of the original surface pattern $\sigma(x,z)$, temperature and
distance from the $y=0$ surface.

In a thin film, creation of truly three dimensional, complex
morphologies between the two
surfaces can be achieved by using only one-dimensional surface
patterns. As an example we choose a simple sinusoidal
pattern on each of the $y=\pm L$ surfaces, rotated 90 degrees 
with respect to one another: $\sigma^- = \cos(qx+qz)$ and
$\sigma^+ =\cos(qx-qz)$. In
Fig.~12(b) the resulting morphology in the $y=0$ mid-plane is
shown and is a superposition of the two surface patterns. Because $y=0$
is a symmetric plane, the pattern has a square symmetry. More complex
patterns can be created at different $y$ planes.

In Sec. \ref{1d} we showed that the $q=0$ mode of the surface pattern
is the slowest decaying mode, resulting in a lamellar layering parallel to the
surface as $y\rightarrow\infty$, no matter what the surface pattern
is. We demonstrate this in Fig.~13, where in (a) we choose a simple
one-dimensional structure in the shape of a stripe of width $d_0$.
Inside the stripe of width $d_0$, 
$\sigma(x,z)=0.5$ while outside it, the surface area is neutral,
$\sigma=0$. Thus, the B-polymer is preferentially adsorbed onto the
stripe. The order parameter contour plot in the $x-y$ plane is shown in (b).
It can be seen that the ``surface disturbance'' is enclosed with
alternating lamellae.  The distorted lamellae close to the $y=0$
surface appear curved, and slowly fade away as the distance from the
surface is increased.  

A different scenario is presented in Fig.~14,
where inside the stripe of thickness $d_0$, $\sigma=0.5$ as above, but
outside the stripe the 
surface is not neutral: $\sigma=-0.5$. We find that 
the adsorption on the surface
is quite different than in Fig.~13. Far from the
stripe, the A-polymer is adsorbed onto the surface and induces stacking
of the BCP in a direction parallel to the surface. Close to the surface 
perturbation (the stripe) the behavior is altered
as the lamellae are strongly deformed in order to optimize their
local interaction with the surface stripe.


\section{Conclusions}\label{conclusions}

We have employed a simple Ginzburg-Landau expansion of the BCP free energy
to study analytically the confinement effects of block copolymers
between two patterned surfaces as well as the interfacial behavior of
a BCP close to a 
patterned surface. Our approach consists of finding the governing
equation for a presumably small perturbation to the bulk order
parameter, by retaining second-order terms in the free energy. This
approach can be justified in the vicinity of the critical point.
Above
the ODT it gives rise to a simple linear equation with fixed
coefficients \cite{Fredrickson87,P-Muthu97}.
A generalization to two-dimensional surface patterns
is presented, where a complete spatial description of
the polymer concentration is given in terms of an arbitrary surface
pattern. However, this approach applies to systems below the ODT as
well, where a linearization is to be taken around an ordered phase
\cite{T-AEPL00,tobepub}.

The assumption that the surface interactions are strictly local means
that the length scales of the polymer morphology are determined by
bulk properties. Moreover, each of the surface $q$-modes in
$\sigma(x)=\sum_q\sigma_q\cos qx$ gives rise to a corresponding mode
$f_q\cos qx$ in the local polymer concentration $\phi(x,y)$.
This ``response'' mode is
characterized by a single wavevector $k_q$. The wavevector $k_q$ is
determined by $\chi$ and the surface wavenumber $q$. In Fig.~1 we show
the dependence of $k_q$ on these parameters. The high $q$-modes of 
the surface pattern $\sigma(x,z)$ decay more rapidly than
those of low $q$. For high $q$-modes of characteristic length scale much
smaller than the polymer chains $d_0$, the BCP melt cannot follow the surface
modulations, and feels just the average of those modulations (which is
zero for $q>0$). This dependence of $k_q$ on $q$ and $\chi$ is very similar to
the results found by Petera and Muthukumar \cite{P-Muthu97} using a
different free-energy functional. 

Moreover, we generalized surface patterns to any
two-dimensional patterns as can be seen in Fig.~10. Even within 
a mode decoupled (linear response) theory,
many interesting effects follow for a single surface as well as for films
confined between two surfaces. Tuning a few surface parameters can lead to
controlled micro-structures of the BCP film.

For a BCP melt in contact with a homogeneous surface, a decaying lamellar
order appears. The phase $\alpha_0$ of these sinusoidally damped
oscillations obeys $\tan\alpha_0=q_0^2/\sqrt{\tau/h}$, and it is
independent of the surface pattern amplitude $\sigma_0$
\cite{B-F-SJPII97}. For high temperatures all $q$-modes have the same
phase $\alpha_q=0$. As $\chi\rightarrow\chi_c$, the $q=q_0$ retains this
value, while higher $q$-modes tend to $-\pi/2$, and lower $q$ tend to
$\pi/2$.  At the same limit the $q=q_0$ mode gets critical first, with
a diverging amplitude.

Our
expressions for the spatial dependence of the order parameter on a
general patterned surface gives a complete description of the system,
and allows for the calculation of free  energy, pressure, etc. It may also
help in tuning the required distance between the two surfaces in
various applications. Using a strong enough surface field or fixing
the conditions close to the ODT, one can hope to transform a pattern
from one surface to the other surface. We also demonstrate in Fig.~12
how the superposition of simple 
one dimensional patterns can bring about a three dimensional
behavior in a thin film system. A desired complex phase can then be
achieved by tuning the Flory parameter and the relevant distances.

Possible extension to this work will be to calculate the phase diagram
of the $L_{\perp}$ phase (a confined lamellar phase where lamellae are
perpendicular to the confining surfaces) vs. the $L_{\parallel}$ phase
(where the lamellae are parallel to the surfaces), by calculating the
surface contribution to the bulk free energy in $F$ (\ref{F}). In the
weak segregation limit this contribution is important and may lead to
a completely different diagram than that of the strong segregation
regime.

\acknowledgements We would like to thank S. Herminghaus, G. Krausch,
M. Muthukumar, R. Netz, G. Reiter, T. Russell, M.  Schick and U.
Steiner for useful discussions.  Partial support from the U.S.-Israel
Binational Foundation (B.S.F.) under grant No. 98-00429 and the Israel
Science Foundation founded by the Israel Academy of Sciences and
Humanities --- centers of Excellence Program is gratefully
acknowledged.

\newpage

\begin{itemize}
  
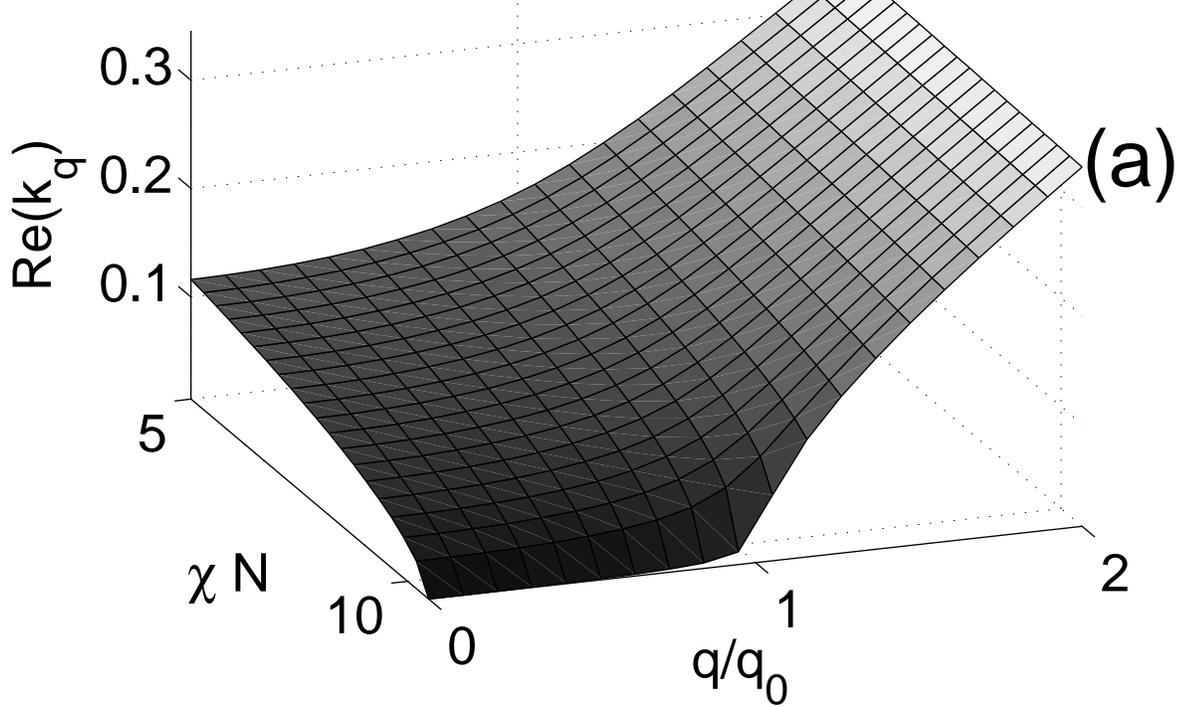
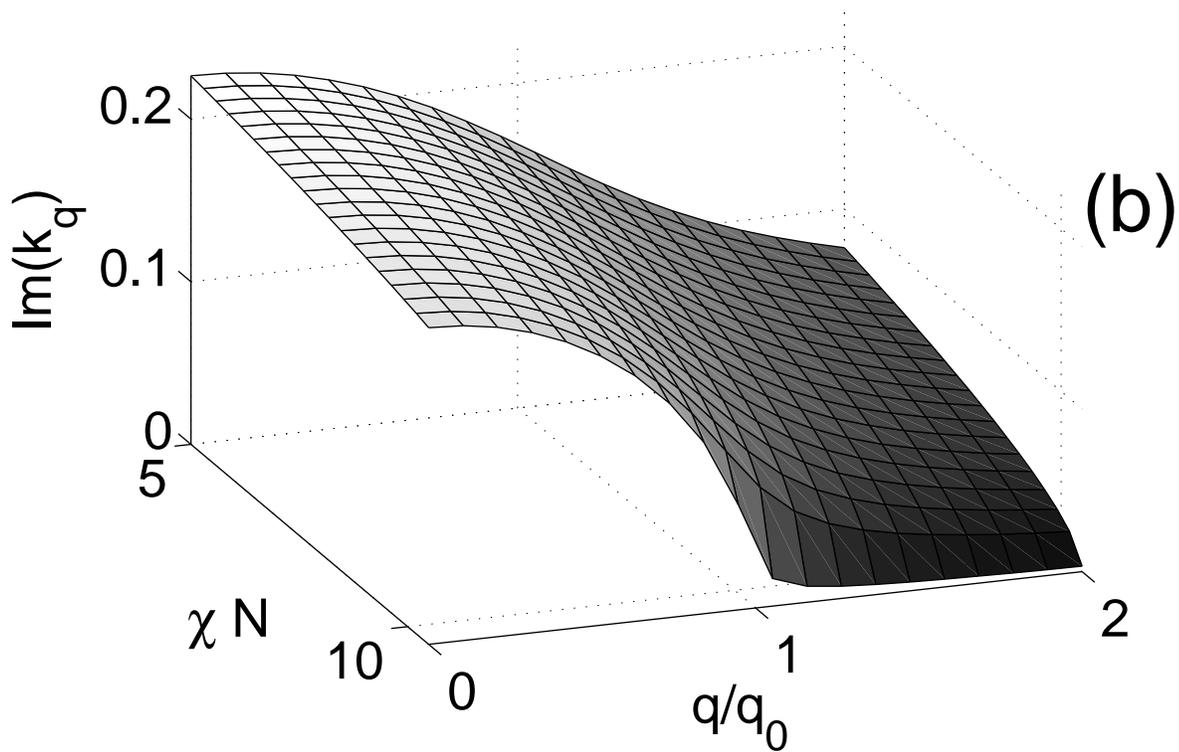
\item{\bf Fig.~1:} The real (a) and imaginary (b) parts of the
  wavevector $k_q$ as a function of the modulation $q$--mode and the
  Flory parameter $N\chi$. For values of $N\chi$ close to
  its critical value,
  $N\chi_c=10.495$, and for small $q$, ${\rm Re}(k_q)$ is small.
  As $q$ increases ${\rm Re}(k_q)$ starts to increase
  rapidly and ${\rm Im}(k_q)$ decreases.
  The value and magnitude of this sharp change in $k_q$ are
  determined by the proximity to ODT. Farther from the critical point
  (smaller $\chi<\chi_c$) the variation of $k_q$ with $q$
  are smoothed out.

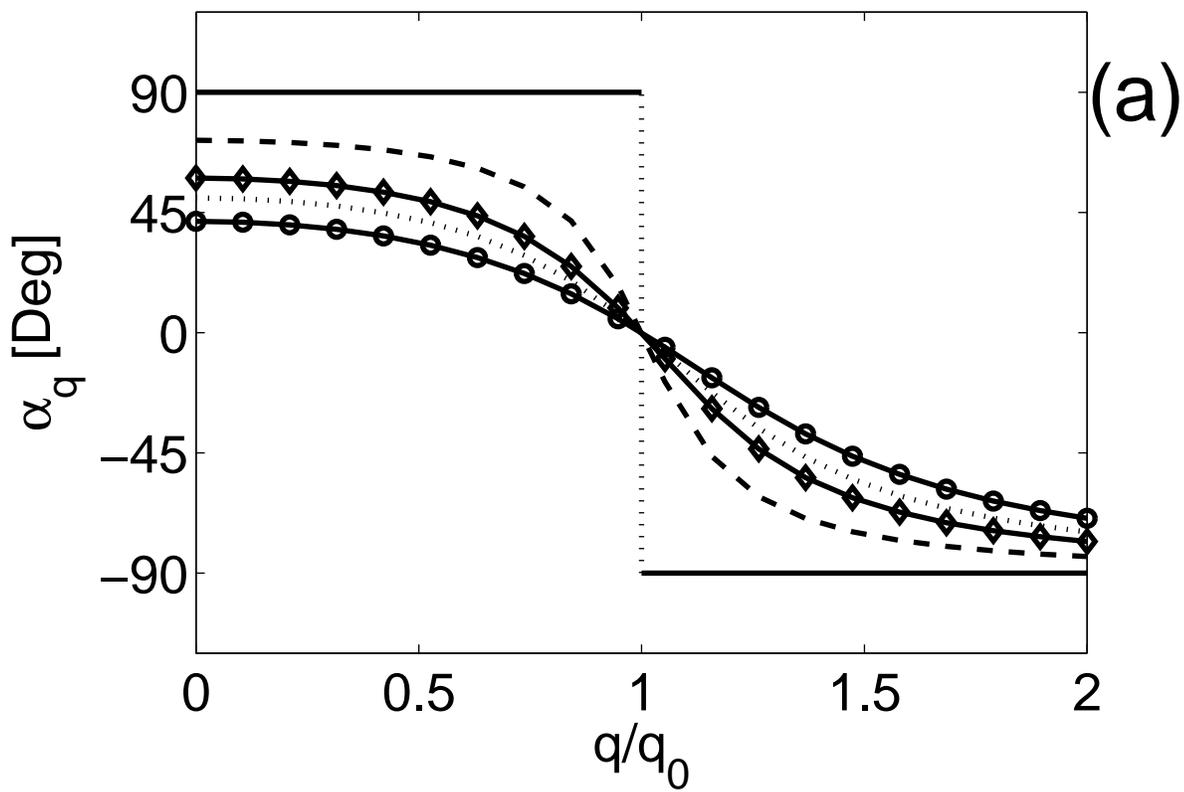
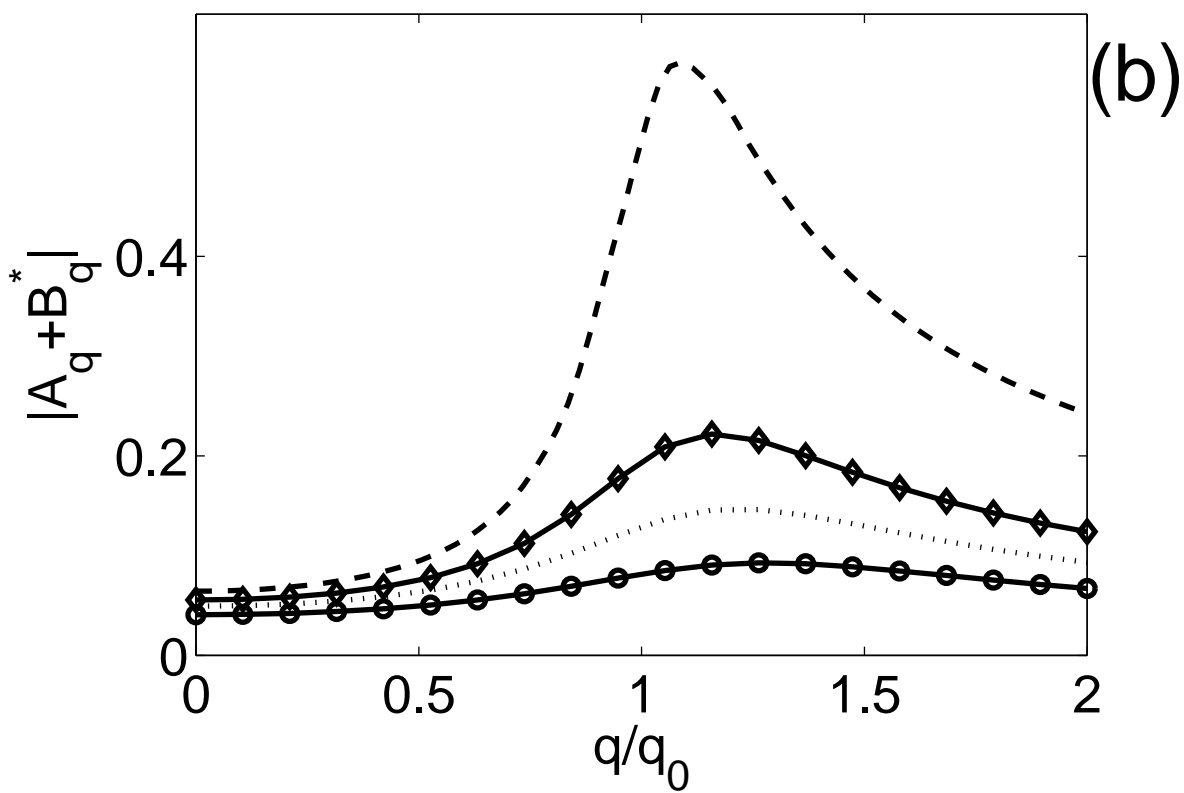
\item{\bf Fig.~2:} (a) A plot of the phase angle $\alpha_q$ from
  Eq. (\ref{alpha}), as a 
  function of the surface $q$-mode, for different Flory parameters
  $N\chi$. Circles, dotted, diamond and dashed lines
  correspond to $N\chi=6.9$, $8.5$, $9.3$ and $10.2$, respectively.
  The solid line is for $N\chi=N\chi_c$. Far from the ODT point (high
  temperatures, $\chi\ll\chi_c$), all $q$-modes have phases equal to
  zero, creating a 
  quarter-lamella region of adsorption near the surface.  In the
  opposite limit, i.e. when $\chi\lesssim\chi_c$, $q$-modes with
  $q<q_0$ have $\alpha_q\rightarrow\pi/2$, while the $q$-modes with
  $q>q_0$ have $\alpha_q\rightarrow-\pi/2$.
  
  In (b) are shown the surface amplitudes $|A_q+B_q^*|$ from Eq.
  (\ref{alpha}), as a function of the $q$-mode, for the same
  series of $\chi$ values as in (a). As $\chi\rightarrow\chi_c$, the
  $q=q_0$ mode goes critical first, with a diverging amplitude.

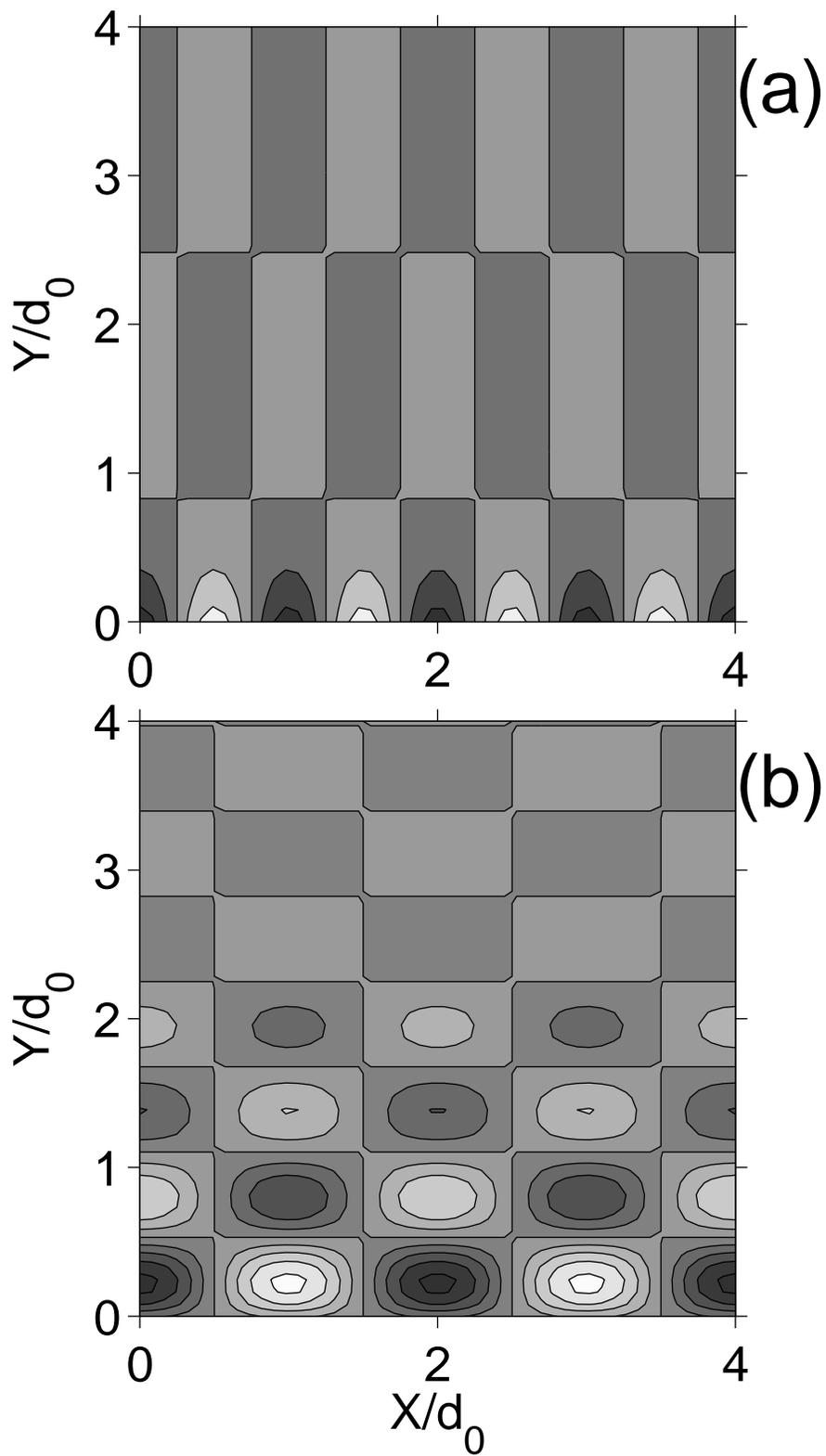
\item{\bf Fig.~3:} A contour plot of the BCP order
  parameter $\phi(x,y)$, where the surface pattern (bottom line, $y=0$)
  contains only one
  mode: $\sigma(x)=\sigma_q\cos(qx)$, with $\sigma_q=1$.
  A-rich regions are black while B-rich are white. In (a), $q=q_0$ while in
  (b) $q=0.5q_0$. The
  decay length $\xi$ is smaller in (a) than in (b) 
  because the surface $q$ mode is larger.
  The Flory parameter was set to $N\chi=10.4<N\chi_c$.
  
\item{\bf Fig.~4:} A contour plot of the BCP order parameter close to
  the critical point ($N\chi=10.4$). The surface
  pattern at $y=0$ is
  $\sigma(x)=\sigma_0+\sigma_q\cos(qx)$, where $\sigma_0$ is the
  average preference and $q=\frac23 q_0$ is the modulation $q$-mode,
  with amplitudes $\sigma_0=\sigma_q=0.1$. The 
  $q=0$ surface mode has a longer range effect than the
  $q>0$ surface mode, and induces parallel lamellar arrangement
  farther away from the surface. At yet larger distances the order
  parameter decays to its bulk $\phi=0$ value.
  
\item{\bf Fig.~5:} A sketch of a thin-film system confined between two
  surfaces.
 The
  $y$ axis is perpendicular to the two parallel surfaces, located at $y=\pm
  L$. The $z$ axis is out of the plane of the paper.

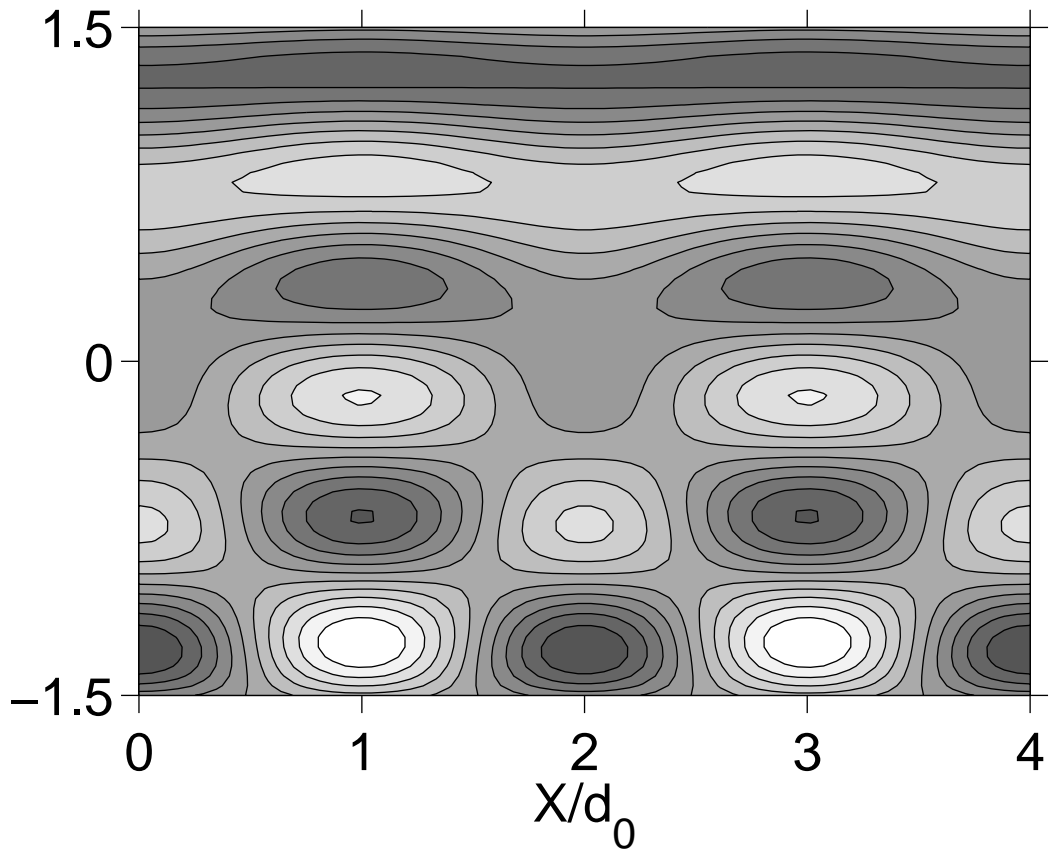
\item{\bf Fig.~6:} Polymer order parameter for a system of homogeneous
  interactions $\sigma^+=1$ at $y=L=1.5d_0$ surface and modulated
  interactions at the opposite surface $\sigma^-=\cos(qx)$, $y=-L$. The
  surface modulation wavenumber was chosen to be $q=0.5 q_0$. As
  expected, the B polymer (shown in black) is preferentially attracted
  to the upper surface, while the bottom surface exhibits modulated
  adsorption pattern. This pattern propagates to the top surface.  The
  Flory parameter was chosen $N\chi=10.4$.

\item{\bf Fig.~7:} The two amplitude functions $f_0(y)$ (dashed line)
  and $f_q(y)$ 
  (solid line) from Fig.~6 plotted against $y/d_0$.  $f_0$ is negative
  at $y=L$ (top uniform surface of Fig.~6), and $f_q$ is negative at the
  opposite modulated surface, $y=-L$. Notice that 
  the pattern at $y=-L$ induces order at the vicinity of the
  other surface, as $f_q(L)\neq 0$, although $\sigma^+=const.$
  
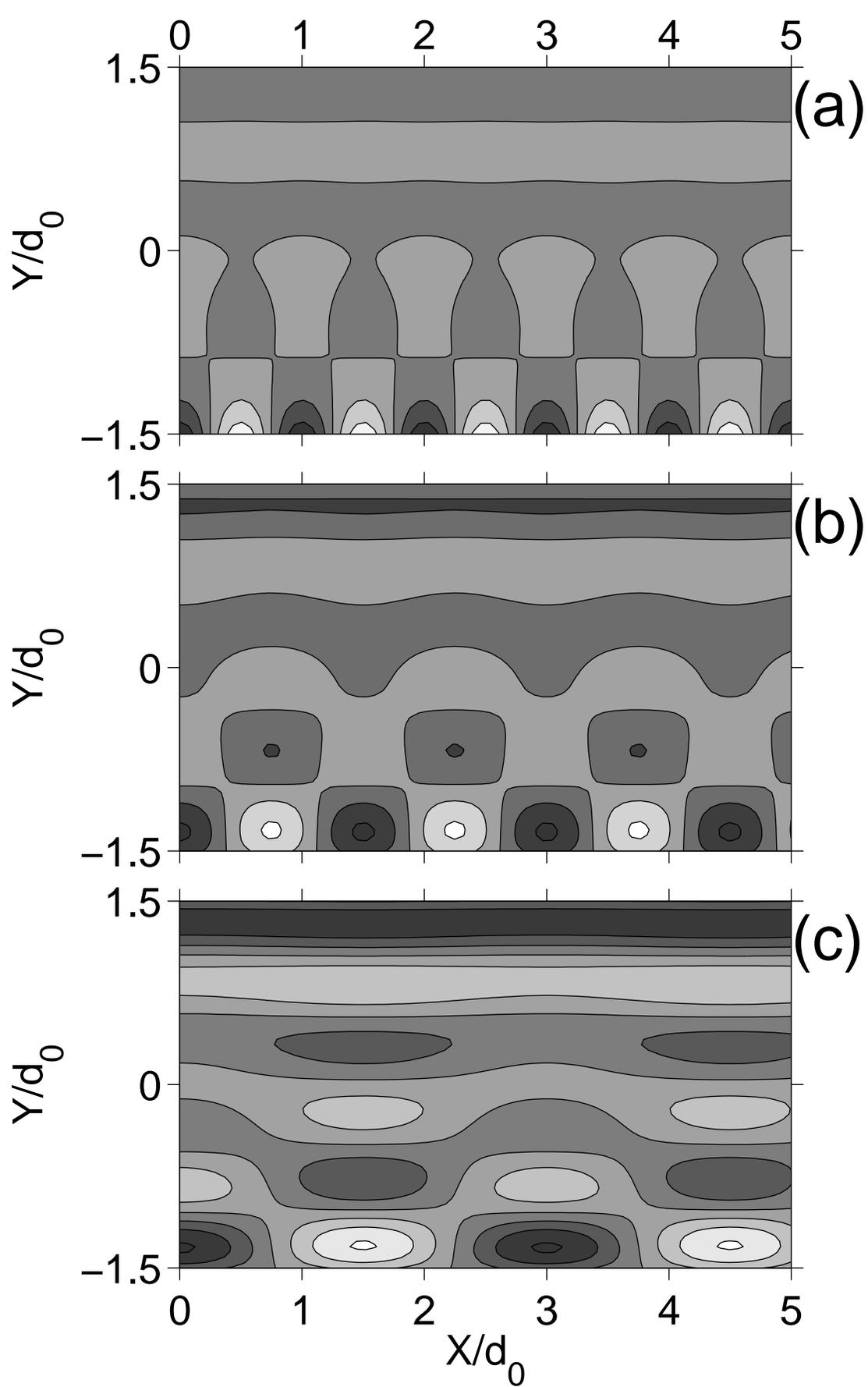
\item{\bf Fig.~8:} A system of modulated surface 
  $\sigma^-=\cos(qx)$ at $y=-L=-1.5d_0$ and of uniform 
  $\sigma^+=1$ at the opposite surface, $y=L=1.5d_0$, for a series of different
  values of $q/q_0$. The effect of changing the repeat period $q$ is
  clearly seen when $q/q_0$ varies: $1$ in (a),
  $2/3$ in (b), $1/3$ in (c). In all cases, $N\chi=10.2$.

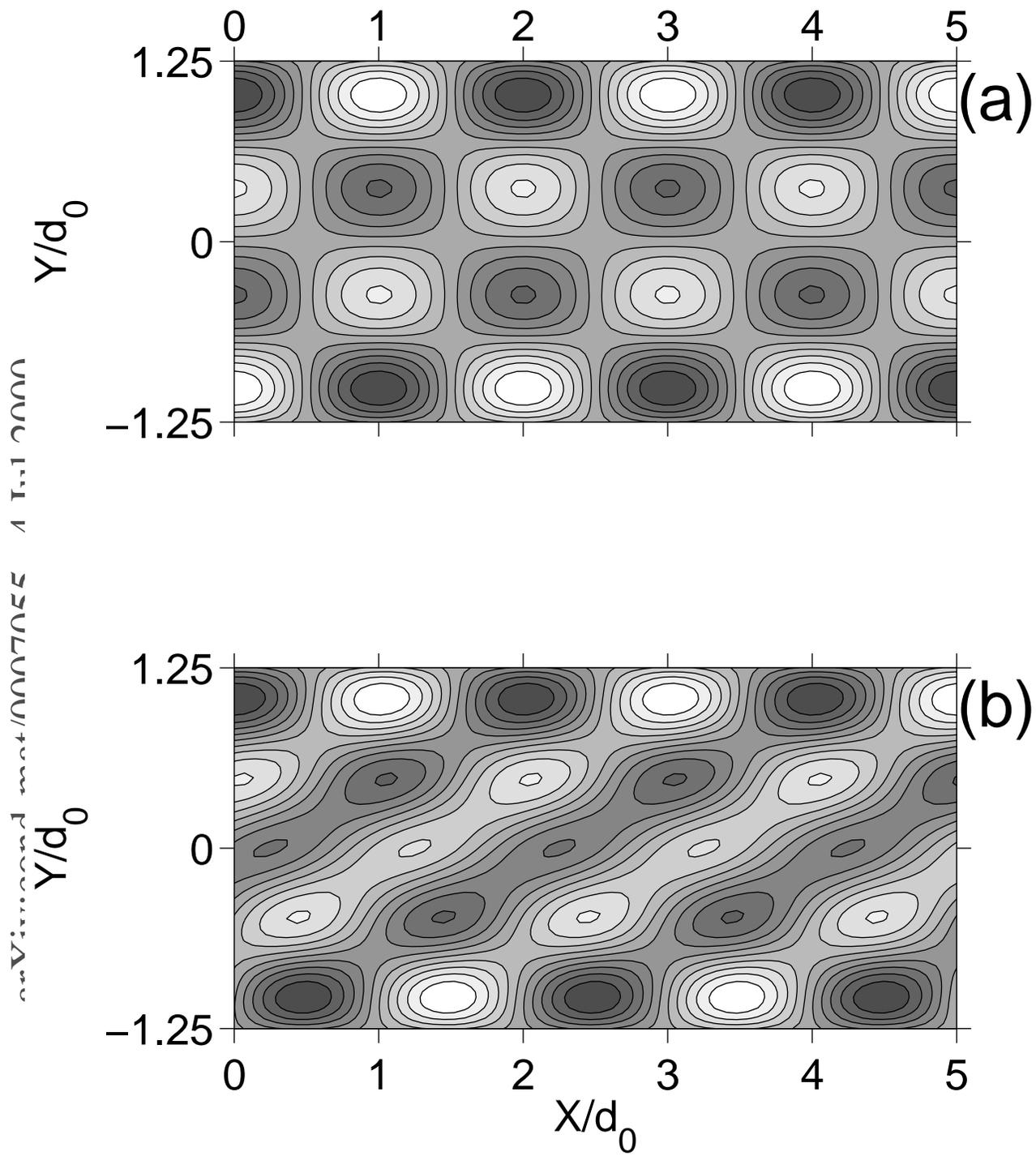
\item{\bf Fig.~9:} The crystalline-like checkerboard character of polymer
  order parameter. In (a) two patterned surfaces in phase with one
  another, but opposite in sign:
  $\sigma^+=-\sigma^-=\sigma_q\cos(\frac12 q_0x)$. In (b) the patterns
  are $\pi/4$ out of phase: $\sigma^+=\sigma_q\cos(\frac12 q_0x)$,
  $\sigma^-=\sigma_q\sin(\frac12 q_0x)$.  
  The amplitude is
  $\sigma_q=0.2$, $N\chi=10.4$, and the top and bottom surfaces are located at
  $y=\pm 1.25d_0$.
  
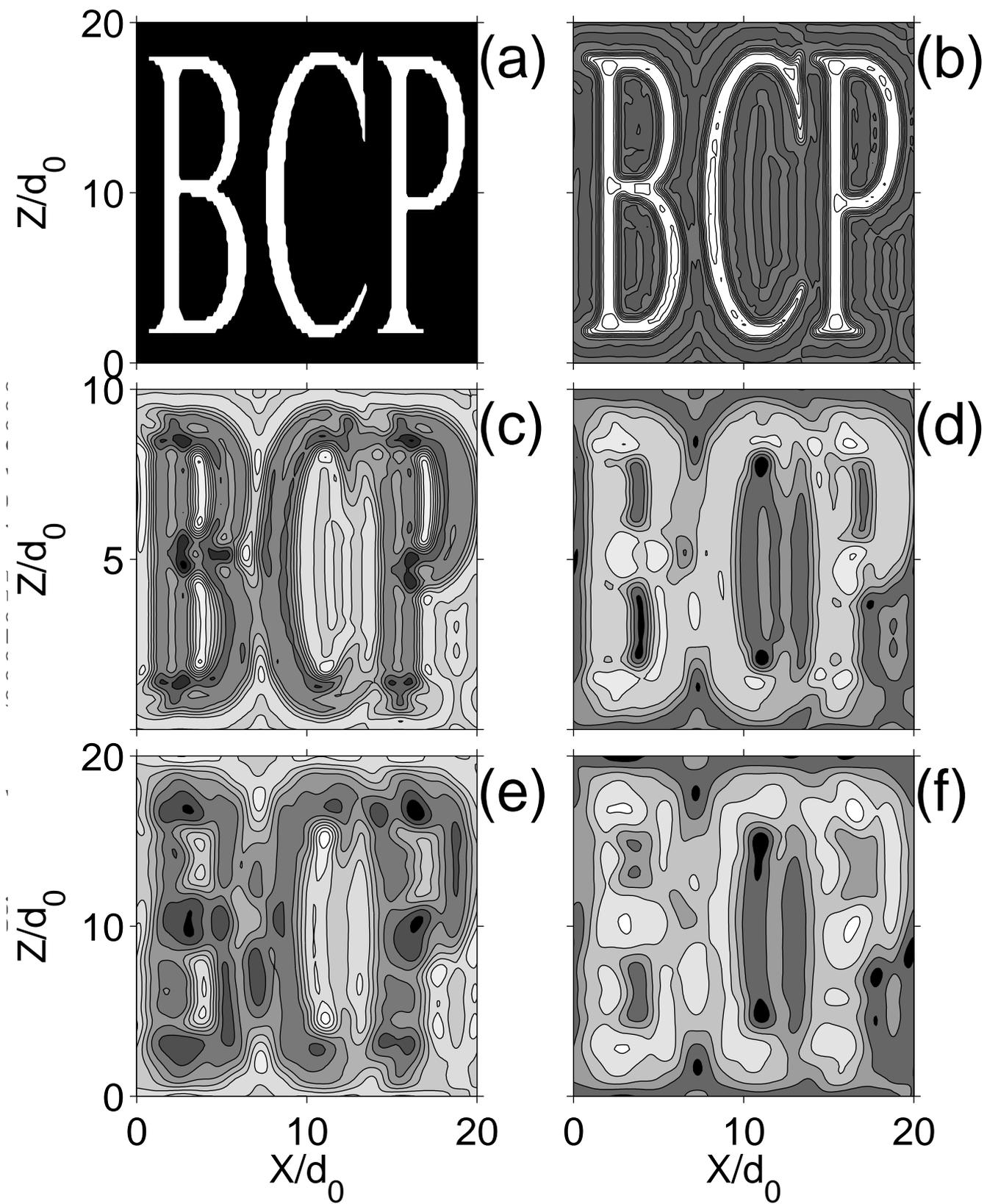
\item{\bf Fig.~10:} Propagation of surface order into the bulk. (a) is
  the original chemical pattern (the letters ``BCP'')
  at the $y=0$ surface, whose size is
  $20\times 20$ in units of $d_0$. White corresponds to A-block
  preferring regions. A sequence of contour plots for $y=0.5$, $2d_0$,
  $3.5d_0$, $5d_0$ and $6.5d_0$ are shown in (b), (c), (d), (e) and
  (f), respectively.  The original pattern is gradually fading (small
  features, high $q$-modes first) as $y$ is increased, until it is
  completely washed out. For $y\approx (n+\frac12)d_0$ with $n$ integer,
  there is an inversion of the original pattern, as the A
  (white) and B block (black) are interchanged relatively to the
  original pattern. The Flory parameter is taken as $N\chi=9.5$.
  
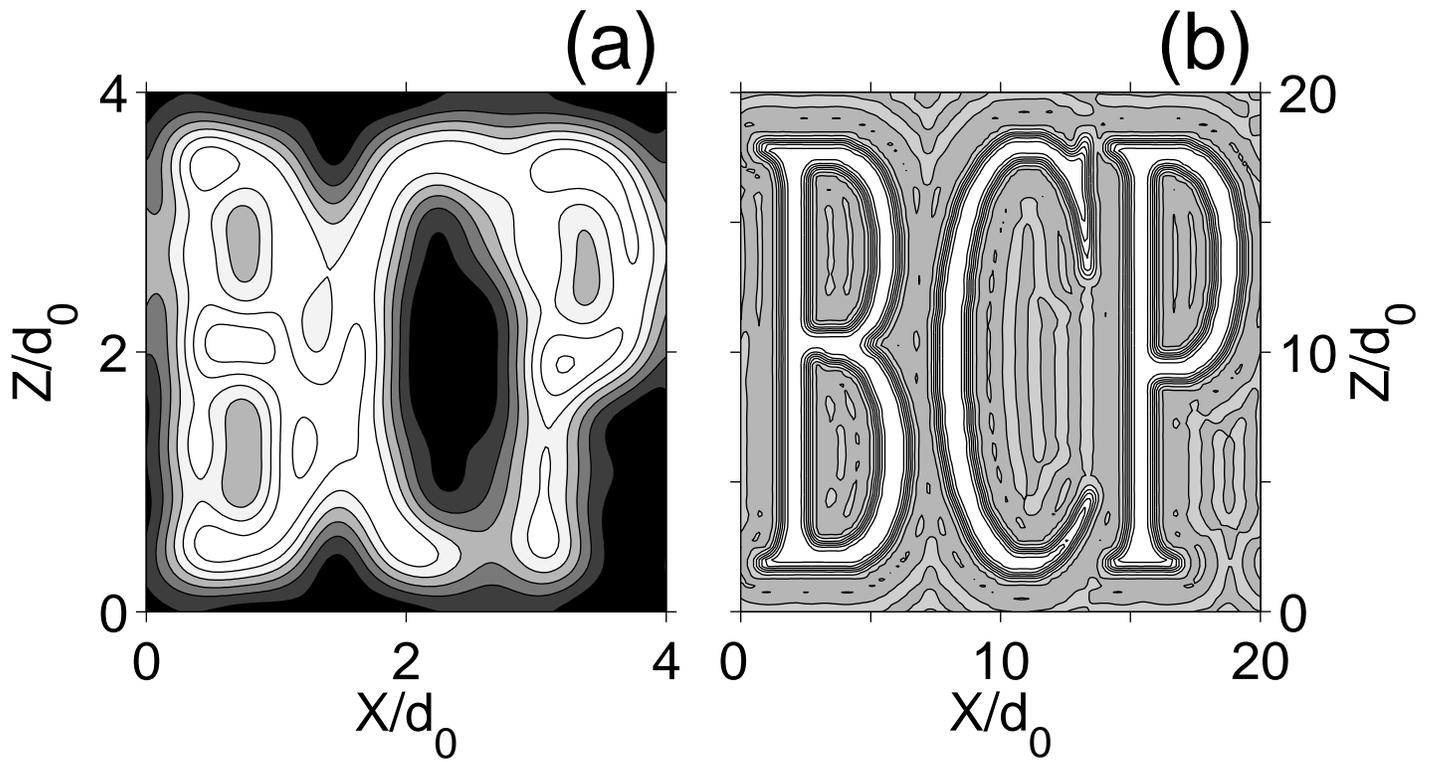
\item{\bf Fig.~11:} Contour plots as in Fig.~10, but in (a) the
  surface pattern is reduced to smaller size of about 
  $4d_0\times 4d_0$, while in (b) size is as in Fig.~10 but the
  temperature is higher, $N\chi=9$. The lamellar features along
  the letter are less prominent than in Fig.~10 (b). Note that in (b) the
  bulk ordering cannot tightly follow the surface pattern when the
  pattern size becomes comparable to $1d_0$, as in (a).

\item{\bf Fig.~12:}
  Creation of a complex three dimensional morphology by superposition
  of two one-dimensional surface patterns:
  $\sigma^-=\cos(qx+qz)$ and $\sigma^+=\cos(qx-qz)$. 
  Shown is the thin film BCP
  morphology, where (a) is the surface pattern at the $y=-L=-d_0$ surface
  and (c) is the pattern at $y=L=d_0$. A contour plot of the order
  parameter in the mid-plane, $y=0$, 
  is shown in (b), depicting A-rich and
  B-rich regions with square symmetry. The Flory parameter is taken as $N\chi=9$
  and $q=q_0/3$.

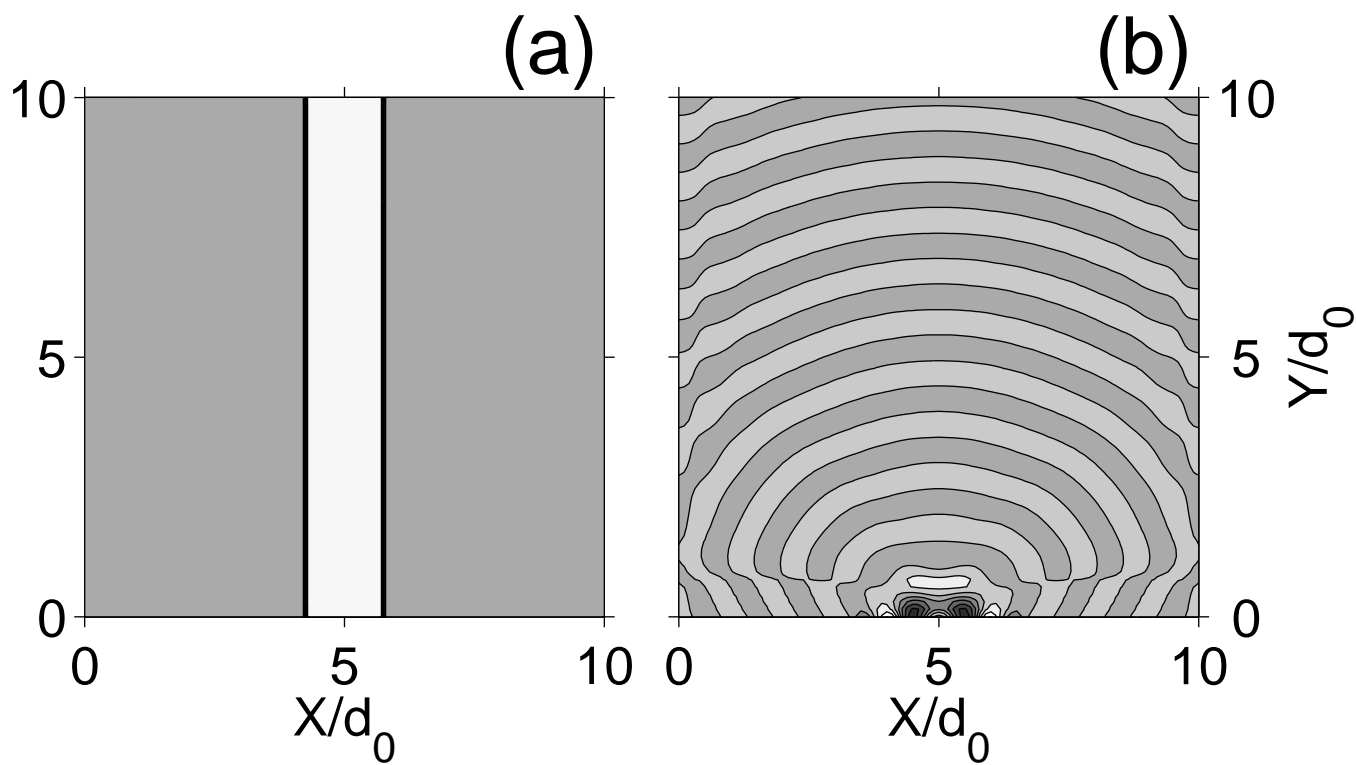
\item{\bf Fig.~13:} Appearance of curved lamellae as a result of a
  one-dimensional surface pattern along the $z$ surface axis. 
  In (a) is surface stripe
  is shown in the $x-z$ plane. The (white) stripe has a surface field of
  $\sigma=0.5$ inducing preferential adsorption of the B-polymer.
  The rest of the surface (denoted by a grey color outside the stripe) 
  has $\sigma=0$ and is indifferent to A/B adsorption.
   (b) is a contour plot in the $x-y$ plane, depicting curved lamellae
  surrounding the ``disturbance'' at the middle.  As the distance from
  the stripe is increased more than $10d_0$ shown here, the lamellae
  gradually fade away. The Flory parameter is taken to be $N\chi=10$.
  
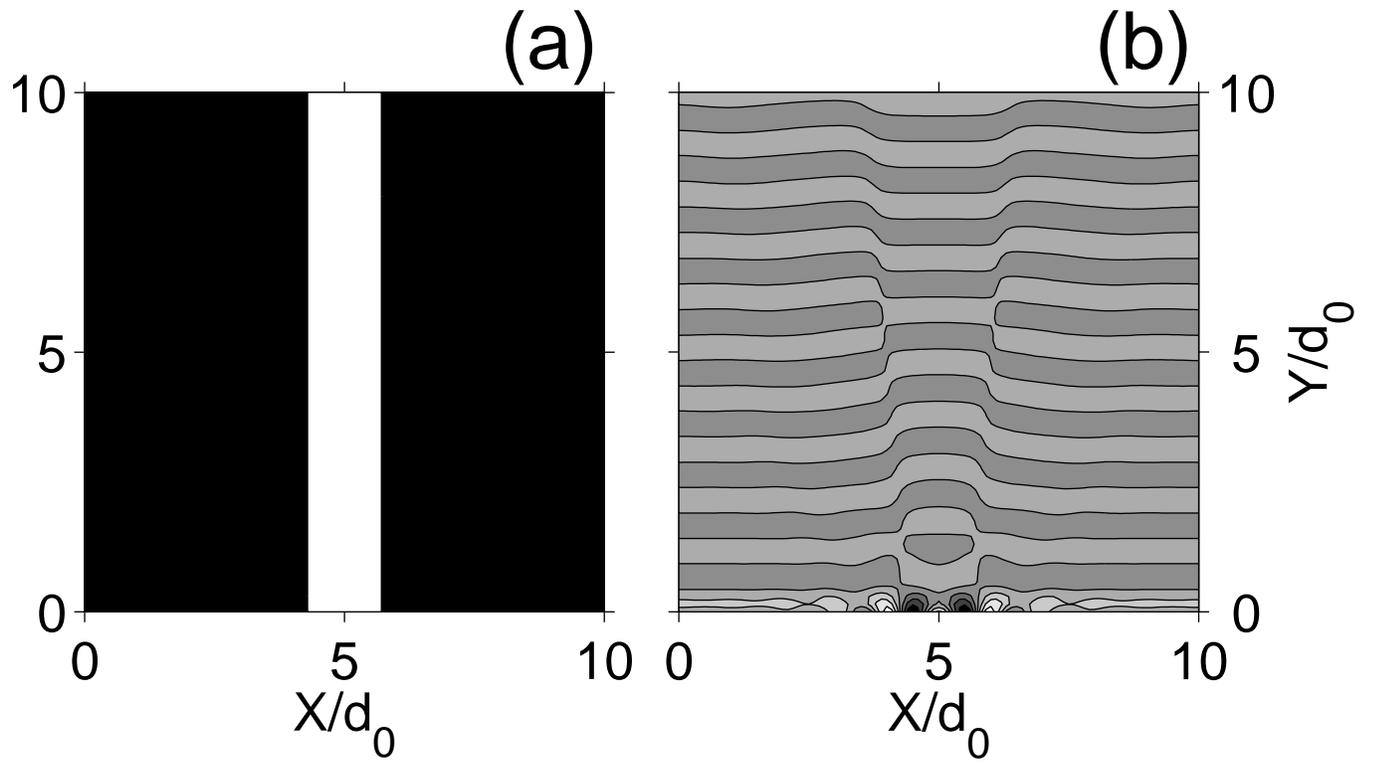
\item{\bf Fig.~14:} Same as in Fig.~13, but with $\sigma=0.5$ inside
  the stripe (white), while the rest of the surface (black) has
  $\sigma=-0.5$. In (a) the $x-z$ surface is shown while in
  (b) the contour plots are shown in the $x-y$ plane. Far from the
  stripe the B-polymer (in black) is adsorbed to the surface, and
  overall a lamellar morphology parallel to the surface is seen. Close
  to the stripe disturbance these lamellae are modified, distorted
  locally by the 
  presence of the stripe.

\end{itemize}

\end{multicols}


\begin{thebibliography}{99}
\bibitem{B-F90} Bates F. S.; Fredrickson G. H. {\it Annu. Rev. Phys.
    Chem.}  {\bf 1990}, {\it 41}, 525.
  
\bibitem{O-K86} Ohta K.; Kawasaki K. {\it Macromolecules} {\bf 1986},
  {\it 19}, 2621.
  
\bibitem{Leibler80} Leibler, L. {\it Macromolecules} {\bf 1980}, {\it
    13}, 1602.

  
\bibitem{F-H87} Fredrickson G. H.; Helfand E. {\it J. Chem. Phys.}
  {\bf 1987}, {\it 87}, 697.
  
\bibitem{M-B96} Matsen M. W.; Bates F. {\it Macromolecules} {\bf
    1996}, {\it 29}, 7641.
  
\bibitem{M-SSCIENCE99} Walheim S.; Sch$\ddot{a}$ffer E.; Mlynek J.;
  Steiner U. {\it Science} {\bf 1999}, {\it 283}, 520.
  
\bibitem{fink98} Fink Y.; Winn J. N.; Fan S.; Chen C.; Michel J.;
  Joannopoulos J.  D.; Thomas E. L. {\it Science} {\bf 1998}, {\it
    282}, 1679.
  
\bibitem{Fredrickson87} Fredrickson G. H. {\it Macromolecules} {\bf
    1987}, {\it 20}, 2535.
  
\bibitem{T-F92} Tang H.; Freed K. F. {\it J. Chem. Phys.} {\bf 1992},
  {\it 97}, 4496.
  
\bibitem{T-R-M94} Turner M. S.; Rubinstein M. R.; Marques C. M.  {\it
    Macromolecules} {\bf 1994}, {\it 27}, 4986.
  
\bibitem{shull92} Shull K. R. {\it Macromolecules} {\bf 1992}, {\it
    25}, 2122.
  
\bibitem{matsenJCP97} Matsen M. W. {\it J. Chem. Phys.} {\bf 1997},
  {\it 106}, 7781.
  
\bibitem{G-M-B00} Geisinger T.; Mueller M.; Binder K.  {\it J. Chem.
    Phys.} {\bf 2000}, {\it 111}, 5241.
  
\bibitem{P-Muthu97} Petera D.; Muthukumar M. {\it J. Chem. Phys.}
  {\bf 1997}, {\it 107}, 9640.
  
\bibitem{P-Muthu98} Petera D.; Muthukumar M. {\it J. Chem. Phys.}
  {\bf 1998}, {\it 109}, 5101.

\bibitem{Binder_Fredrickson89} Fredrickson, G.H.; Binder K. 
{\it J. Chem. Phys.} {\bf 1989 }, {\it 91}, 7265. 

\bibitem{B-F-SJPII97} Binder K.; Frisch H. L.; Stepanow S. {\it J.
    Phys. II} {\bf 1997}, {\it 7}, 1353.

  
\bibitem{brazovskii} Brazovskii S. A., {\it Sov. Phys. JETP} {\bf
    1975}, {\it 41}, 85.
  
\bibitem{T-A-S00} Tsori Y; Andelman D.; Schick M. {\it Phys. Rev. E.}
  {\bf 2000}, {\it 61}, 2848.
  
\bibitem{G-S90} Gompper G.; Schick M. {\it Phys. Rev. Lett.} {\bf
    1990}, {\it 65}, 1116.

  
\bibitem{N-A-SPRL97} Netz R. R.; Andelman D.; Schick M. {\it Phys.
    Rev.  Lett.} {\bf 1997}, {\it 79}, 1058.

  
\bibitem{A-B-J87} Andelman D.; Brochard F.; Joanny J.-F. {\it J. Chem.
    Phys.} {\bf 1987}, {\it 86}, 3673.
  
\bibitem{G-D82} Garel T.; Doniach S. {\it Phys. Rev. B} {\bf 1982},
  {\it 26}, 325.
  
\bibitem{L-RPRL96} Kellogg G. J.; Walton D. G.; Mayes A. M.; Lambooy
  P.; Russell T. P.; Gallagher P. D.; Satija S. K. {\it Phys.  Rev.
    Lett.} {\bf 1996}, {\it 76}, 2503.
  
\bibitem{M-RPRL97} Mansky P.; Russell T. P.; Hawker C. J.; Mayes J.;
  Cook D. C.; Satija S. K. {\it Phys. Rev. Lett.} {\bf 1997}, {\it
    79}, 237.

  
\bibitem{K-Muthu99} Kielhorn L.; Muthukumar M. {\it J. Chem. Phys.}
  {\bf 1997}, {\it 111}, 2259.
  
\bibitem{V-N-A-S98} Villain-Guillot S.; Netz R. R.; Andelman D.;
  Schick M.  {\it Physica A} {\bf 1998}, {\it 249}, 285.
  Villain-Guillot S.; Andelman D. {\it Euro. Phys. J. B} {\bf 1998},
  {\it 4}, 95.
  
\bibitem{mukamelPRE00} Jacobs A. E.; Mukamel D.; Allender D. W.  {\it
    Phys. Rev. E} {\bf 2000}, {\it 61} 2753.
  
\bibitem{G-TMAC94} Gido S. P.; Thomas E. L.  {\it Macromolecules} {\bf
    1994}, {\it 27}, 6137.

  
\bibitem{T-AEPL00} Tsori Y.; Andelman D. submitted to {\it EuroPhys.
    Lett}.

  
\bibitem{tobepub} Tsori Y.; Andelman D. to be published.


\end{thebibliography}
\end{document}